\documentclass[apj]{emulateapj}
\def\gtorder{\mathrel{\raise.3ex\hbox{$>$}\mkern-14mu
                \lower0.6ex\hbox{$\sim$}}}
\def\ltorder{\mathrel{\raise.3ex\hbox{$<$}\mkern-14mu
                \lower0.6ex\hbox{$\sim$}}}
\slugcomment{}
\shorttitle{Ion-by-ion Cooling Efficiencies}
\shortauthors{O. Gnat \& G. Ferland}

\begin{document}
\title{Ion-by-Ion Cooling Efficiencies}
\vspace{1cm}
\author{Orly Gnat\altaffilmark{1,2} and Gary J. Ferland\altaffilmark{3}}
\altaffiltext{1}{Theoretical Astrophysics, California Institute of Technology, 
                 MC 350-17, Pasadena, CA 91125, USA.}
\altaffiltext{2}{Chandra Fellow}
\altaffiltext{3}{Department of Physics and Astronomy, University of Kentucky, 
                 Lexington, KY 40506, USA.}
\email{orlyg@tapir.caltech.edu}

\begin{abstract}
We present ion-by-ion cooling efficiencies for low-density gas. We use Cloudy 
(ver.~08.00) to estimate the cooling efficiencies for each ion of the first $30$
elements (H$-$Zn) individually. We present results for gas temperatures between 
$10^4$ and $10^8$~K, assuming low densities and optically thin conditions. 
When nonequilibrium ionization plays a significant role
the ionization states deviate
from those that obtain in collisional ionization equilibrium (CIE),
and the local cooling efficiency at any given temperature 
depends on specific non-equilibrium ion fractions. The results presented here
allow for an efficient estimate of the total cooling efficiency for any ionic
composition. We also list the elemental cooling efficiencies assuming CIE 
conditions. These can be used to construct CIE cooling efficiencies for non-solar
abundance ratios, or to estimate the cooling due to elements not 
included in any nonequilibrium computation. All the computational results are
listed in convenient online tables.
\end{abstract}


\section{Introduction}
\label{introduction}

The radiative cooling efficiencies of hot ($10^4-10^8$~K) low density gas are
important quantities in the study of the diffuse interstellar and intergalactic
medium. They determine the thermal and dynamical properties and evolution in
a variety of astrophysical systems, ranging from local interstellar clouds 
to shocks in intergalactic filaments.

Computations of hot gas cooling efficiencies assuming collisional ionization
equilibrium (CIE) have been studied extensively (House~1964; 
Tucker \& Gould~1966; Allen \& Dupree~1969; Cox \& Tucker~1969;
Jordan~1969; Raymond et al.~1976; Shull \& van Steenberg~1982;
Gaetz \& Salpeter~1983; Arnaud \& Rothenflug~1985; Boehringer \& Hensler~1989;
Sutherland \& Dopita~1993; Landi \& Landini~1999; Benjamin et al.~2001).
These CIE cooling efficiencies depend only on the gas temperature and
metallicity. However, there are many cases in which CIE conditions do not apply.
For example, nonequilibrium ionization is bound to occur when an initially hot 
gas cools radiatively below $\sim10^6$~K (Kafatos~1973; Shapiro \& Moore~1976;
Edgar \& Chevalier~1986; Schmutzler \& Tscharnuter~1993; Sutherland \& 
Dopita~1993, and Smith et al.~1996; Gnat \& Sternberg~2007). Below
this temperature, cooling becomes rapid compared to electron-ion
recombinations, and the gas at any temperature tends to remain
``overionized''compared to gas in CIE. In conduction fronts surrounding
evaporating clouds (e.g.,~Borkowski et al.~1990; Gnat et al.~2010), 
non-equilibrium ionization occurs when the ionization time is long compared 
to the rate of temperature change. In this case, the gas tends to remain 
underionized compared to CIE. Non-equilibrium ionization also plays a role in 
fast radiative shock waves (e.g.,~Allen et al.~2009; Gnat \& Sternberg~2009)
and in turbulent mixing layers (e.g.,~Slavin et al.~1993).

When departures from CIE are significant, the cooling efficiencies are no longer
a function of just the gas temperature and metallicity, but instead depend on
the specific time-dependent ion fractions. The use of convenient tables with
known CIE cooling efficiencies must be replaced with a detailed computation of
the non-equilibrium cooling, taking into account all the relevant microphysical
processes which include numerous emission lines, thermal bremsstrahlung, and 
ionization and recombination cooling. This both requires the collection of 
a large set of atomic data for all the relevant processes, and is computationally complex compared
with using look-up tables.

Recently, first attempts have been made at including nonequilibrium ionization
physics in large scale hydrodynamical simulations, with applications for both
cosmological metal-absorption through the warm/hot intergalactic medium 
(e.g.,~Cen \& Ostriker~2006; Cen \& Fang~2006) and emission-lines from
galaxy clusters (e.g.,~Akahori \& Yoshikawa~2010). Because of the complexity 
of nonequilibrium cooling, such 
simulation have so far not included self-consistent nonequilibrium cooling
efficiencies. It is simpler to include nonequilibrium cooling in 
hydrodynamical simulations by using look-up tables for the nonequilibrium
cooling rate (e.g.,~Oppenheimer \& Dav{\'e}~2009).

In this paper, we present ion-by-ion cooling efficiencies. We list the cooling
efficiencies for each ion of the first $30$ elements (H$-$Zn) individually.
We present results for gas temperatures between $10^4$ and $10^8$~K, assuming
optically-thin, low-density conditions. The sum of ionic cooling efficiencies, 
weighted by
the nonequilibrium ion-densities, then provides an efficient-to-compute and
self-consistent nonequilibrium cooling efficiency. We also list the elemental
cooling efficiencies assuming CIE conditions. These can be used to
construct CIE cooling efficiencies for non-solar abundance ratios, or 
to estimate the remaining cooling due to elements not included in any 
time-dependent computation.

This paper is {\it not} a new calculation of the cooling functions
(c.f. Schure et al.~2009). We do not present any new atomic data.
Instead, we list the current cooling functions included in Cloudy
(ver. 08.00) in convenient online tables that are useful in any
numerical computation in which the ion abundances are not in
photoionization or collisional equilibrium.
It is the first time that the individual ionic cooling efficiencies are 
listed in an accessible format. As we describe below, the new 
framework that we present here will be periodically maintained
and updated, as improved atomic data becomes available.

The outline of this paper is as follows. In Section~2, we describe the 
computational method. In Section~3, we present the results for the ion-by-ion
cooling efficiencies, and for the element-by-element CIE cooling efficiencies.
We summarize in Section~4.

\section{Method}
\label{method}

We used Cloudy (ver.~08.00, Ferland et al.~1998) to compute the 
ion-by-ion cooling efficiencies of every ion of the first thirty 
elements (H$-$Zn). All the cooling processes considered by Cloudy are 
described in detail in Osterbrock \& Ferland (2006), and include 
collisional excitations followed by line emission, recombinations with 
ions, collisional ionizations, and thermal bremsstrahlung\footnote{Cloudy
is available at http://www.nublado.org/. The code documentation, including a 
full description of all cooling processes is available at 
http://viewvc.nublado.org/index.cgi/tags/release/c08.00/- docs/?root=cloudy
(and will be described in Ferland et al. 2011, in preparation).
To get the full list of references to the atomic data in this version, 
see instructions in
section 13.5 of Cloudy's third volume of documentation (Hazy3$\_$08.pdf).
}.
The electron cooling efficiency includes the removal of electron {\it
kinetic} energy via recombinations with ions, collisional ionizations, collisional
excitations followed by prompt line emissions, and thermal bremsstrahlung.
Cloudy does not include the ionization potential energies as part of the
total internal energy, but instead follows the loss and gain of the electron
kinetic energy only. Therefore, in the definition of the cooling
(see Osterbrock \& Ferland 2006; Gnat \& Sternberg 2007) the ionization 
potential energy that is released
as recombination radiation does not appear. Only the kinetic energy of the
recombining electrons contributes to the cooling efficiency. On the other 
hand, kinetic energy removed via collisional ionization is included in the 
cooling. If ionization potential energy is considered as part of the total internal
energy, then collisional ionization does not lead to a net energy loss,
since the kinetic energy removed is merely stored as potential energy.
Either way of accounting for the energy losses leads to the same
{\it net} (i.e. cooling  minus heating) cooling efficiency.

For each ionization state $i$ of each element $E$, we constructed a
series of models for different gas temperature between $10^4$ and $10^8$~K.
Each model includes only hydrogen and the element $E$. 
The abundances of all other elements are set to zero.
We set an electron density $n_e=1$~cm$^{-3}$, 
regardless of composition and ionization state. We define the ionization 
states of the element $E$ so that the fractional abundance of the species
$E_i$ is $1$, and the abundances of all other ions are $0$. 
We further set the abundance of element $E$ to be $10^{15}$ larger than
that of Hydrogen, so that $n_{\rm H}=10^{-15}$~cm$^{-3}$ and 
$n(E_i)=1$~cm$^{-3}$. Effectively, each such model contains only 
the species $E_i$ and free electrons at the specified electron 
temperature.\footnote{We verify that the contribution of Hydrogen to each 
model is negligible, by changing the abundance of the relevant species 
from $10^{15}$ times $n_{\rm H}$ to $10^{10}$ times $n_{\rm H}$, and 
verifying that the results remain unaltered.}

For each element, we also compute the cooling efficiencies assuming CIE
ion fractions.
In these models, we set the abundance of the element $E$ to be $10^{15}$ 
larger than that of Hydrogen (so that again $n_{\rm H}=10^{-15}$~cm$^{-3}$ and 
$n(E)=1$~cm$^{-3}$), and we force an electron density $n_e=1$~cm$^{-3}$. 
We allow Cloudy to compute the CIE ion fractions $x(E_i)$. For each 
element $E$, the ion fractions $x(E_i) = n(E_i)/n(E)$, must at all
times satisfy,
\begin{equation}
\sum_i x(E_i) = 1\;,
\end{equation}
where $n(E_i)$ is the density (cm$^{-3}$) of ions in ionization stage $i$
of element $E$, $n(E)=n_H A_E$, $n_H$ is the total hydrogen density, and $A_E$
is the abundance of element $E$ relative to hydrogen. These models yield the 
total cooling efficiency due to the CIE ion distribution of element $E$.

The tables presented in this paper provide easy access to different components 
of the total cooling function computed in version 08.00 of Cloudy.  
The cooling function included in Cloudy is constantly being
updated as improved atomic data become available.
We will update this table, keeping its current format, to provide ready access 
to these future calculations.  This way, codes that can parse the current tables
can be easily updated as better atomic data become available.

\section{Ion-by-Ion cooling efficiencies}
\label{results}

We have carried out computations of the cooling efficiencies 
$\Lambda_{e,{\rm ion}}(T)$ for each ion of the first $30$ elements,
H$-$Zn (with atomic numbers $1-30$), as a function of temperature.
The results are listed in tabular form in Table~\ref{res}. The full
table is available at http://wise-obs.tau.ac.il/$\sim$orlyg/ion$\textunderscore$by$\textunderscore$ion/, 
and is divided into lettered parts A-AD, as is outlined in Table~\ref{guide}.
For each element $E$ with atomic number $Z$, the first column in 
Table~\ref{res} lists the temperature, and the next $Z+1$ columns
list the cooling efficiencies, $\Lambda_{e,{\rm ion}}$ 
(erg~s$^{-1}$~cm$^3$) for the different ionization states, starting 
with the neutral atom, and ending with the fully stripped ion.
The cooling rate per unit volume due to ionization state $i$ of
element $E$, is then given by $n({\rm ion})\,n_e\,\Lambda_{e,{\rm ion}}$ 
(erg~s$^{-1}$~cm$^{-3}$). For example, Table~\ref{res}A shows that
the cooling rate due to neutral hydrogen at a temperature of 
$1.1\times10^4$~K is $1.33\times10^{-23}\,n({\rm H}^0)\,n_e$~erg~s$^{-1}$~cm$^{-3}$
whereas the cooling efficiency due to ionized hydrogen at the same
temperature is $6.47\times10^{-25}\,n({\rm H}^+)\,n_e$~erg~s$^{-1}$~cm$^{-3}$.

\begin{deluxetable}{cccc}
\tablewidth{0pt}
\tablecaption{Ion-by-Ion Cooling Efficiencies\label{res}}
\tablehead{
\colhead{T}&
\colhead{$\Lambda_{e,{\rm H\,I}}($\ion{H}{1}$)$}&
\colhead{$\Lambda_{e,{\rm H\,II}}($\ion{H}{2}$)$}&
\colhead{$\Lambda_{e,{\rm H}}($H at CIE$)$}\\
\colhead{(K)}&
\colhead{ (erg~cm$^3$~s$^{-1}$)}&
\colhead{ (erg~cm$^3$~s$^{-1}$)}&
\colhead{ (erg~cm$^3$~s$^{-1}$)}
}
\startdata
$1.00\times10^{4}$ & $4.59\times10^{-24}$ & $6.26\times10^{-25}$ & $4.58\times10^{-24}$ \\
$1.05\times10^{4}$ & $7.93\times10^{-24}$ & $6.37\times10^{-25}$ & $7.90\times10^{-24}$ \\
$1.10\times10^{4}$ & $1.33\times10^{-23}$ & $6.47\times10^{-25}$ & $1.32\times10^{-23}$ \\
\enddata
\tablecomments{Table~\ref{res} is available in its entirety 
at http://wise-obs.tau.ac.il/$\sim$orlyg/ion$\textunderscore$by$\textunderscore$ion/. 
A portion is shown here for guidance regarding
its form and content. The full table lists the ion-by-ion cooling efficiencies for all
the different ions of the first 30 elements (H$-$Zn), and for the elemental cooling 
efficiencies assuming CIE. For a guide see Table~\ref{guide}.}
\end{deluxetable}

\begin{deluxetable}{lllll}
\tablewidth{0pt}
\tablecaption{Cooling Data and Solar Elemental Abundances\label{guide}}
\tablehead{
\colhead{Z}&
\colhead{Element}&
\colhead{Table}&
\colhead{Abundance (X/H)$_\odot$}&
\colhead{Reference$^{1}$}
}
\startdata
1  & Hydrogen    &  1A  & $1                 $  &    \\
2  & Helium      &  1B  & $8.33\times10^{-2} $  & B  \\
3  & Lithium     &  1C  & $2.04\times10^{-9} $  & C  \\
4  & Beryllium   &  1D  & $2.63\times10^{-11}$  & C  \\
5  & Boron       &  1E  & $6.17\times10^{-10}$  & C  \\
6  & Carbon      &  1F  & $2.45\times10^{-4} $  & A  \\
7  & Nitrogen    &  1G  & $6.03\times10^{-5} $  & A  \\
8  & Oxygen      &  1H  & $4.57\times10^{-4} $  & A  \\
9  & Fluorine    &  1I  & $3.02\times10^{-8} $  & C  \\
10 & Neon        &  1J  & $1.95\times10^{-4} $  & DT \\
11 & Sodium      &  1K  & $2.14\times10^{-6} $  & C  \\
12 & Magnesium   &  1L  & $3.39\times10^{-5} $  & A  \\
13 & Aluminum    &  1M  & $2.95\times10^{-6} $  & C  \\
14 & Silicon     &  1N  & $3.24\times10^{-5} $  & A  \\
15 & Phosphorus  &  1O  & $3.20\times10^{-7} $  & C  \\
16 & Sulfur      &  1P  & $1.38\times10^{-5} $  & A  \\
17 & Chlorine    &  1Q  & $1.91\times10^{-7} $  & C  \\
18 & Argon       &  1R  & $2.51\times10^{-6} $  & C  \\
19 & Potassium   &  1S  & $1.32\times10^{-7} $  & C  \\
20 & Calcium     &  1T  & $2.29\times10^{-6} $  & C  \\
21 & Scandium    &  1U  & $1.48\times10^{-9} $  & C  \\
22 & Titanium    &  1V  & $1.05\times10^{-7} $  & C  \\
23 & Vanadium    &  1W  & $1.08\times10^{-8} $  & C  \\
24 & Chromium    &  1X  & $4.68\times10^{-7} $  & C  \\
25 & Manganese   &  1Y  & $2.88\times10^{-7} $  & C  \\
26 & Iron        &  1Z  & $2.82\times10^{-5} $  & A  \\
27 & Cobalt      &  1AA & $8.32\times10^{-8} $  & C  \\
28 & Nickel      &  1AB & $1.78\times10^{-6} $  & C  \\
29 & Copper      &  1AC & $1.62\times10^{-8} $  & C  \\
30 & Zinc        &  1AD & $3.98\times10^{-8} $  & C  \\
\enddata
\tablecomments{
(1) References: A: Asplund et al~(2005); B: Ballantyne et al.~(2000); 
C: adopted from Cloudy, based on Grevesse \& Sauval~(1998);
DT: Drake \& Testa~(2005).}
\end{deluxetable}

\begin{figure}[!h]
\epsscale{1.2}
\plotone{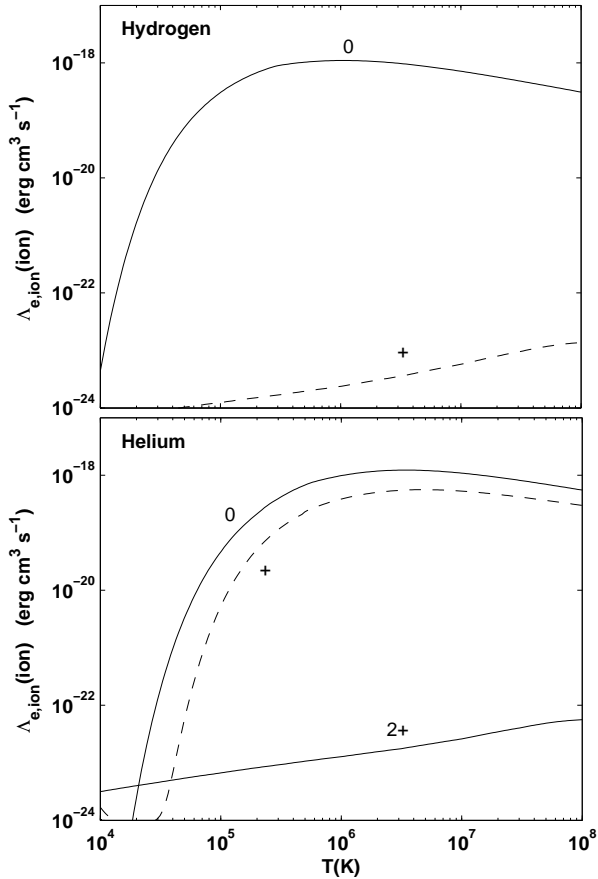}
\caption{Ion-by-ion Cooling efficiencies for hydrogen ions (upper panel)
and for helium ions (lower panel).}
\label{HHe}
\end{figure}

Figure~\ref{HHe} shows the ion-by-ion cooling efficiencies for hydrogen
(upper panel) and for helium (lower panel). This Figure confirms that
if neutral hydrogen exists at high temperatures, it cools
orders of magnitude more efficiently than ionized hydrogen. For example,
at a temperature of $10^6$~K, neutral hydrogen has a cooling efficiency
of $\sim10^{-18}$~erg~cm$^3$~s$^{-1}$, due to collisional ionizations and
Ly$\alpha$ emission. Ionized hydrogen has a cooling efficiency of 
$\sim2\times10^{-24}$~erg~cm$^3$~s$^{-1}$, due to thermal bremsstrahlung
emission.

\begin{figure}[!h]
\epsscale{1.2}
\plotone{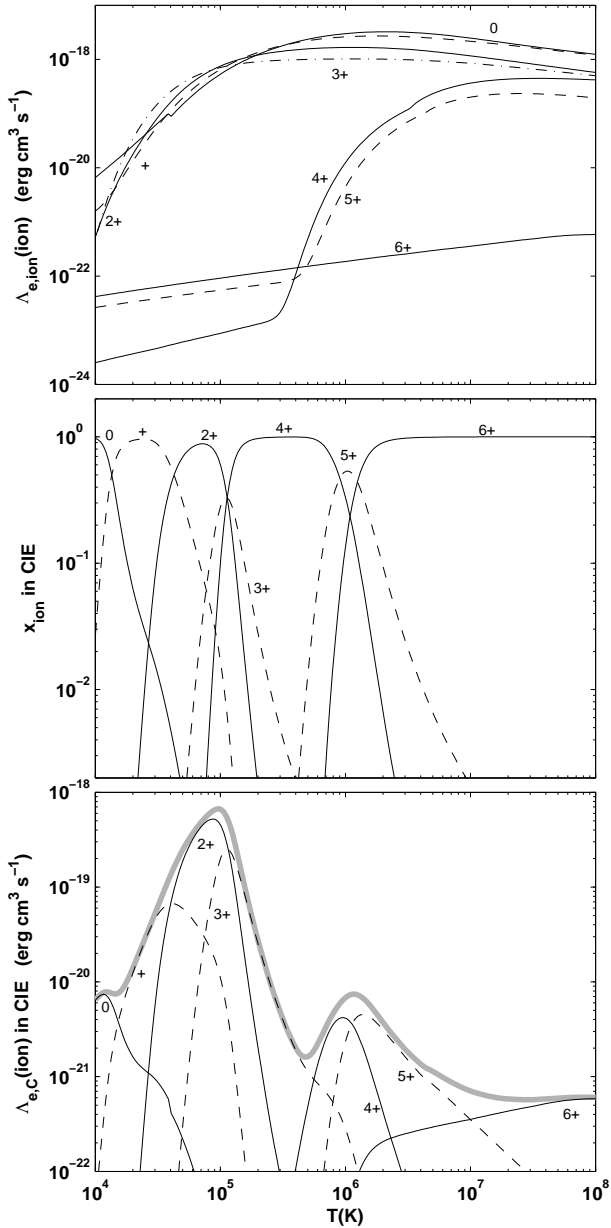}
\caption{Cooling efficiencies of carbon ions. {\it Upper panel}: Ion-by-ion
cooling efficiencies for carbon ions. The cooling rate per unit volume
is given by $n_e n_{\rm ion} \Lambda_{e,{\rm ion}}({\rm ion})$. {\it Middle Panel}:
Carbon CIE ion fractions. {\it Lower Panel}: Carbon CIE cooling efficiencies
for carbon ions, $x_{\rm ion}\Lambda_{e,{\rm ion}}$. The cooling rate per
unit volume is given by $n_e n_{\rm C} \Lambda_{e,{\rm C}}({\rm ion})$. The sum of 
CIE cooling efficiencies by all carbon ions is shown by the thick gray curve.}
\label{C}
\end{figure}

The total cooling due to a specific element depends of the ion abundances.
Figure~\ref{C} shows the cooling efficiencies vs. temperature for the different
carbon ions (see Table~\ref{res}F). The top panel shows
the ion-by-ion cooling efficiencies, 
$\Lambda_{e,{\rm ion}}$. These can be used to compute the total carbon
cooling efficiency for any composition. As an example, in the middle panel
we show the carbon CIE ion fractions. The bottom panel then shows the 
contribution of each ion to the CIE carbon cooling efficiency
$\Lambda_{e,\rm{C}}({\rm ion}) = x_{\rm ion}\Lambda_{e,{\rm ion}}$.
For example, at a given carbon density $n({\rm C})$, the contribution of
C$^{3+}$ to the CIE cooling rate per volume is 
$n({\rm C})\,n_e\,\Lambda_{e,\rm{C}}({\rm C}^{3+})$~erg~s$^{-1}$~cm$^{-3}$.
The sum of CIE cooling efficiencies over all carbon ions is shown by the
thick gray curve in the lower panel. The carbon CIE cooling efficiency has
two peaks.  The first peak, at $6\times10^4-2\times10^5$~K is due to
C$^+$, C$^{2+}$, and C$^{3+}$. These ions are responsible for the familiar
carbon peak in the solar-metallicity CIE cooling curve at 
$\sim10^5$~K (see Section~\ref{CIEs}).
The second peak, at $\sim10^6$~K, is due to C$^{4+}$ and C$^{5+}$, and
is two order of magnitude lower.

\subsection{Element-by-Element CIE cooling}
\label{CIEs}

\begin{figure*}[!h]
\epsscale{1.0}
\plotone{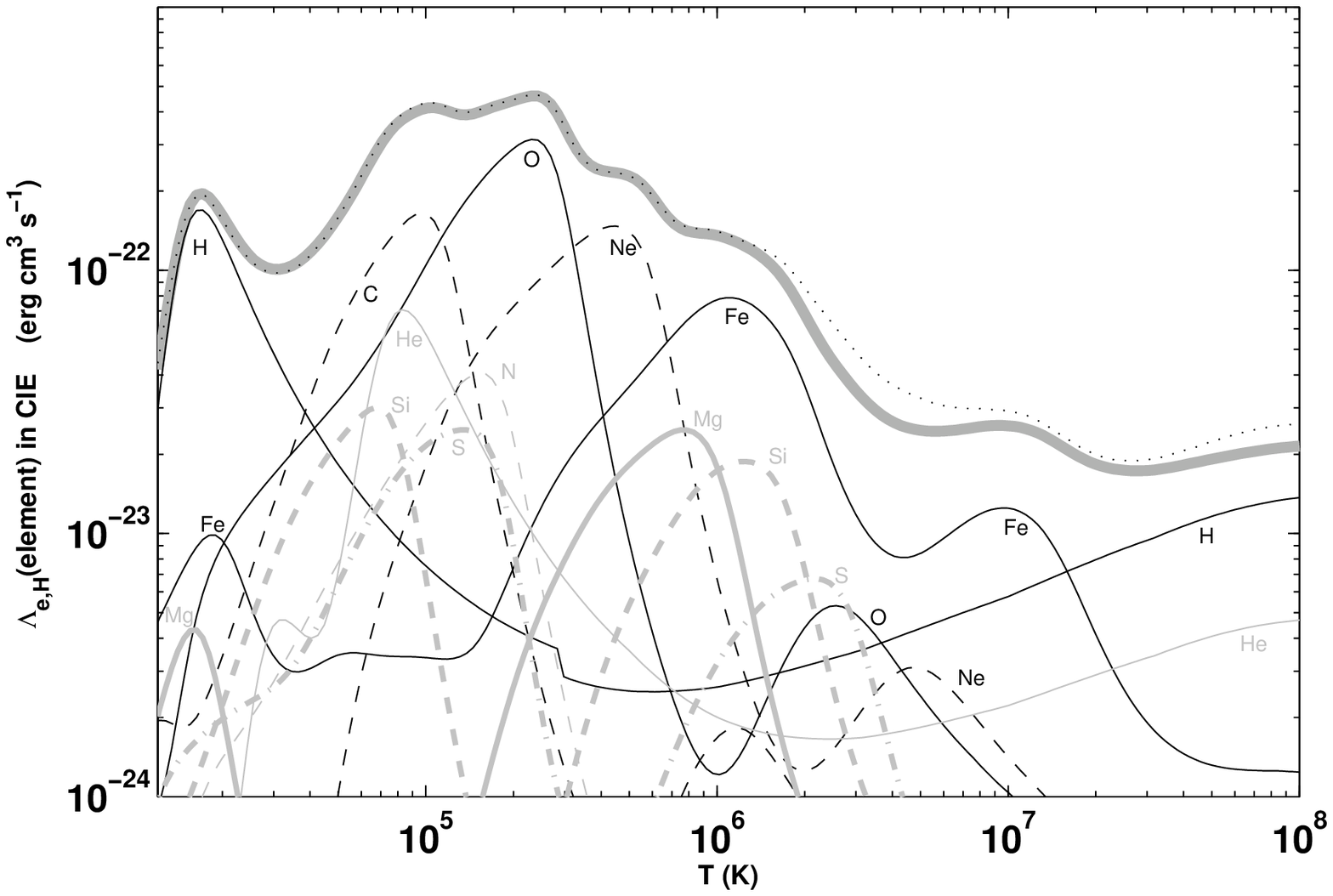}[t]
\caption{Element-by-element cooling efficiencies assuming CIE ion fractions
and solar elemental abundance ratios (see Table~\ref{guide}). The total CIE
cooling efficiency due to all elements is shown by the upper thick gray
curve. The CIE cooling efficiency of Gnat \& Sternberg~(2007), which relied
on Cloudy version~06.02 is shown by the upper dotted line for comparison. 
The differences between the two are due to updated atomic data included in
Cloudy version~08.00.}
\label{CIE}
\end{figure*}

The last column in each part (A-AD) of Table~\ref{res}
lists the total CIE cooling efficiency
of each element as a function of temperature. Figure~\ref{CIE} shows the
CIE cooling efficiencies of the major coolants as a function of 
temperature, assuming a solar metallicity. In making 
Figure~\ref{CIE}, we adopt the
elemental abundances for C, N, O, Mg, Si, S, and Fe reported by Asplund et
al.~(2005) for the photosphere of the Sun, and the enhanced Ne abundance
recommended by Drake \& Testa~(2005). For the other elements we use the
abundances reported by Grevesse \& Sauval~(1998). We list these abundances
in Table~\ref{guide}.

Figure~\ref{CIE} shows the familiar peaks in the CIE cooling efficiency
due to different elements. The low-temperature peak at $\sim2\times10^4$~K
is mainly due to hydrogen Ly$\alpha$ cooling. As the hydrogen neutral fraction
becomes small, the contribution of hydrogen Ly$\alpha$ decreases. This peak is
followed by peaks at $10^5$, $3\times10^5$, $5\times10^5$, and $1.5\times10^6$~K
due, respectively, to contributions of carbon, oxygen, neon, and iron.
A second iron peak can be seen at $\sim10^7$~K. At higher temperatures
cooling is dominated by thermal bremsstrahlung due to fully stripped ions.
Contributions by other cooling elements are also shown in figure~\ref{CIE}.
For example, cooling due to helium peaks at a temperature of $\sim8\times10^4$~K. 
For solar metallicity gas the helium contribution to the total cooling 
is small compared with that of metal-line cooling. However, at subsolar 
metallicities the relative
contribution of helium is larger. Nitrogen, magnesium, silicon, and 
sulfur also contribute to the cooling below $\sim5\times10^6$~K. 
The results for the elemental CIE cooling
efficiencies are in qualitative agreement with previous computations (e.g.,~Sutherland
\& Dopita~1993). Difference in detail are mainly due to differences in assumed
atomic data (see Gnat \& Sternberg~2007), affecting both the CIE ion fractions
and the cooling efficiencies of specific cooling processes.

The upper thick gray curve shows the total contribution of all elements to
the CIE, solar metallicity, cooling efficiency. These results are identical 
to the cooling efficiencies computed by Cloudy (ver. 08.00) 
assuming CIE and a solar composition (including all elements). 
For comparison, the dotted curve shows the CIE cooling
efficiencies of Gnat \& Sternberg~(2007), which were computed using the cooling
function included in Cloudy ver. 06.02. The agreement is excellent for
$T\lesssim2\times10^6$~K, but some differences appear at higher temperatures.
This is due to difference in the input atomic data between the two Cloudy
versions.


\section{Summary}
\label{summary}

In this paper, we present computations of the cooling efficiencies
of each ion of the first 30 elements (hydrogen-zinc) individually.
We use the cooling functions included in Cloudy (ver. 08.00) to
compute the cooling efficiencies as a function of temperature,
between $10^4$ and $10^8$~K, assuming optically thin conditions.

The results are listed in tabular form in Table~\ref{res} 
(Section~\ref{results}), and are available in convenient online
format through the electronic edition of the journal (for a guide
see Table~\ref{guide}). For each ion, we list the cooling 
efficiency $\Lambda_{e,{\rm ion}}(T)$ (erg~cm$^3$~s$^{-1}$)
as a function of temperature. The total cooling rate for any
ionic composition can then be computed by multiplying the ionic
efficiencies by the ion densities,
$n_e \sum_{\rm ion} n({\rm ion}) \Lambda_{e,{\rm ion}}$
(erg~s$^{-1}$~cm$^{-3}$).

As opposed to gas in CIE, for which the cooling efficiencies
depend only on the gas temperature and metallicity, for nonequilibrium
conditions the cooling efficiencies must be evaluated locally
depending on the nonequilibrium ion fractions. 
A self-consistent computation therefore requires the collection of a 
large set of atomic data for all the relevant microphysical cooling processes,
including numerous emission lines, thermal bremsstrahlung, and ionization 
and recombination processes. The results presented in this paper allows for
an efficient estimate of the total cooling efficiency regardless of the
ionization state.

The tables presented here use the current atomic data set within Cloudy
ver. 08.00.  The tables provide a flexible way to access the cooling of
individual species.  The Cloudy atomic database is continuously updated
and new versions of these tables, using the same format, will be created
as the atomic data are improved\footnote{Cloudy provides a means to generate
its current atomic physics bibliography, as described in Section~2 above.}.  
These tables will then provide easy access to future improvements as they 
occur.

The online tables are useful when constructing theoretical models in 
which nonequilibrium ionization plays a significant role, and can be 
used, for example, in models for radiatively cooling gas, conduction 
fronts, fast shock waves, and turbulent mixing layers. They can also 
simplify the inclusion of self-consistent nonequilibrium cooling in
large scale cosmological hydrodynamical simulations.

In Section~\ref{CIEs}, we present the {\it elemental} cooling efficiencies
as a function of temperature, for each of the first 30 elements
(H$-$Zn) assuming CIE conditions. These results can be used to easily
construct CIE cooling efficiencies for non-solar abundance ratios,
as well as to estimate the cooling by elements not included
in any time-dependent, nonequilibrium computations.

\vspace{1cm}
\section*{Acknowledgments}

OG was supported by NASA through Chandra Postdoctoral Fellowship
grant PF8-90053 awarded by the Chandra X-ray Center, which
is operated by the Smithsonian Astrophysical Observatory for NASA 
under contract NAS8-03060.
Partial financial support for GJF's work on this project was provided
by National Science Foundation grants AST 0908877 and AST 0607028, National
Aeronautics and Space Administration grant 07-ATFP07-0124, and HST Theory
grant AR 12125.01.


\end{document}